\documentclass[conference]{IEEEtran}
\IEEEoverridecommandlockouts
\usepackage{cite}
\usepackage{amsmath,amssymb,amsfonts}
\usepackage{algorithmic}
\usepackage{graphicx}
\usepackage{textcomp}
\usepackage{xcolor}
\usepackage{booktabs}
\usepackage{multirow}
\usepackage{arydshln}

\def\BibTeX{{\rm B\kern-.05em{\sc i\kern-.025em b}\kern-.08em
    T\kern-.1667em\lower.1ex\hbox{E}\kern-.125emX}}

\IEEEoverridecommandlockouts
\usepackage{bm}
\usepackage{cite}
\usepackage{amsmath,amssymb,amsfonts}
\usepackage{algorithmic}
\usepackage{graphicx}
\usepackage{threeparttable}
\usepackage{textcomp}
\usepackage{algorithmic}
\usepackage{algorithm}
\usepackage{url}
\usepackage{xcolor} % For coloring text
\usepackage{times}  % for Times font
\usepackage{mathptmx}

\makeatother

\begin{document}

\onecolumn
\vspace{+5cm}
\begin{center}
    \LARGE{\textbf{\textcolor{red}{This paper has been accepted for presentation in the 2025 Amercian Control Conference (ACC).}}}
\end{center}
\twocolumn

\title{\LARGE \bf{Model-Free Generic Robust Control for Servo-Driven Actuation Mechanisms with Layered Insight into Energy Conversions}}
\author{Mehdi Heydari Shahna and Jouni Mattila
\thanks{This work was supported by the Business Finland Partnership Project, `Future All-Electric Rough Terrain Autonomous Mobile Manipulators' under Grant No. 2334/31/2022. (Corresponding author: Mehdi Heydari Shahna.)}
\thanks{The authors are with the Faculty of Engineering and Natural Sciences, Tampere University, Tampere, Finland
        (e-mail: mehdi.heydarishahna@tuni.fi; jouni.mattila@tuni.fi).}
}

\maketitle

\begin{abstract}
To advance theoretical solutions and address limitations in modeling complex servo-driven actuation systems experiencing high non-linearity and load disturbances, this paper aims to design a practical model-free generic robust control (GRC) framework for these mechanisms. This framework is intended to be applicable across all actuator systems encompassing electrical, hydraulic, or pneumatic servomechanisms, while also functioning within complex interactions among dynamic components and adhering to control input constraints. In this respect, the state-space model of actuator systems is decomposed into smaller subsystems that incorporate the first principle equation of actuator motion dynamics and interactive energy conversion equations. This decomposition operates under the assumption that the comprehensive model of the servo-driven actuator system and energy conversion, uncertainties, load disturbances, and their bounds are unknown. Then, the GRC employs subsystem-based adaptive control strategies for each state-variant subsystem separately. Despite control input constraints and the unknown interactive system model, the GRC-applied actuator mechanism ensures uniform exponential stability and robustness in tracking desired motions. It features straightforward implementation, experimentally evaluated by applying it to two industrial applications.
\end{abstract}

\section{Introduction}
\label{sec:introduction}
\begin{figure}[h!]
    \hspace*{-0.0cm} % Adjust the value as needed
    \centering
    \scalebox{0.7}{\includegraphics[trim={0cm 0.0cm 0.0cm 0cm},clip,width=\columnwidth]{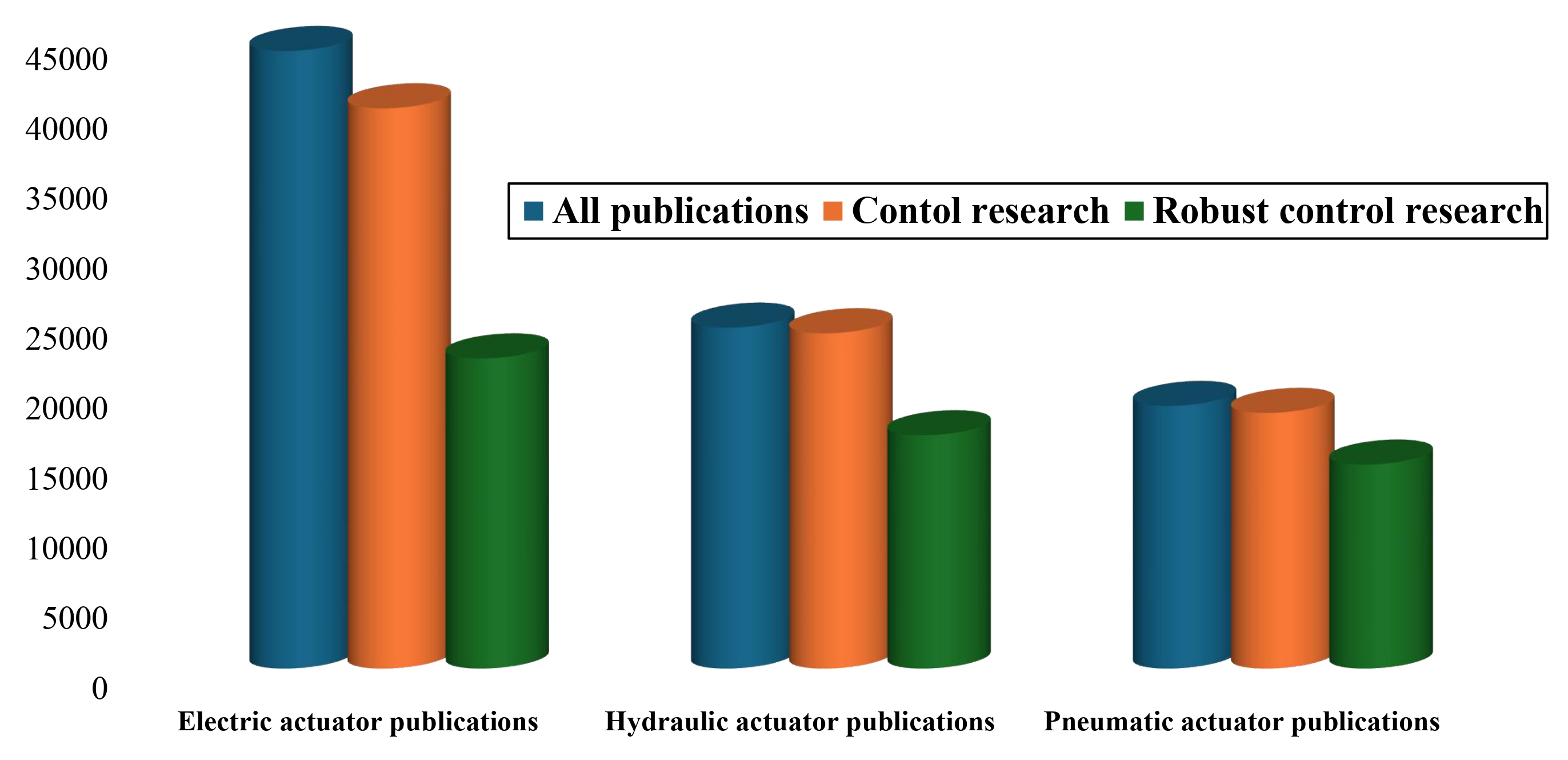}}
    \caption{Research distribution in EDAs, HDAs, and PDAs since 2020.}
    \label{fig1}
\end{figure}

Despite significant advances in control theory and the design of intelligent controls in industries, proportional–integral–derivative (PID) control is widely used in industrial applications for managing position and force due to its straightforward and model-free implementation. However, PID control might be adequate only for systems primarily characterized by second-order dynamics. \cite{zhao2017pid, zhang2019theory}. The effectiveness of PID control is often restricted in higher-order systems, leading to more complex, multi-modal oscillatory behavior, poor performance, or even potential instability \cite{aastrom2006advanced}. Primary actuation systems typically exhibit multi-component dynamics and inherent nonlinearities, highly complicating the accurate control of position and force \cite{guo2017saturated, sciavicco2012modelling, mattila2017survey}. The most commonly employed servomechanism actuators in industrial applications include hydraulically driven actuators (HDAs), pneumatically driven actuators (PDAs), and electrically driven actuators (EDAs). HDAs offer considerable power-to-weight ratios, rapid response times, and the ability to handle heavy loads \cite{yao2017active}, while PDAs are favored for their affordability and cleanliness \cite{ren2019adaptive}. However, EDAs are becoming increasingly favored for their high performance potential, primarily outshining hydraulic actuators in terms of efficiency and reduced maintenance requirements \cite{heydari2024robust}. These actuation systems function by converting the input from electrical, pneumatic, or hydraulic servomechanisms into mechanical energy. Such conversions enable the controlled movement necessary for the system’s operation. Thus, the control system design for the actuation mechanisms is divided into two main stages \cite{heydari2024robust, yang2017position, ren2019adaptive}. The outer stage describes the first principle of mechanical dynamics, calculating the required torque/force to ensure the actuator's position or velocity aligns with the desired values. The inner stage, called energy conversion (or power) equations, describes how energy in electrical, hydraulic, or pneumatic form is converted into mechanical energy. In these systems, the torque or force is directly related to the oil pressure/flow in HDAs \cite{yang2017position, yao2017active}, air pressure/flow in PDAs \cite{ren2019adaptive}, or stator currents in EDAs \cite{heydari2024robust}. Nonetheless, controlling these highly complex and multi-component systems effectively poses difficulties due to their naturally nonlinear characteristics. Uncertainties and disturbances in actuators driven by servomechanisms can stem from fluctuations in supply pressure or voltage \cite{yang2016disturbance}, variations in control servo dynamics, changes in the bulk modulus, sensor inaccuracies, and load effects, among other factors \cite{yao2017active, ren2019adaptive, heydari2024robust}. Fig. \ref{fig1} shows a comprehensive literature search using the keywords `electric actuators,' `hydraulic actuators,' and `pneumatic actuators' through Google Scholar and other academic databases. This search illustrates the distribution of publications across these most common types of actuators since 2020, underscoring the significance of designing robust control systems for these actuators, given that they operate across diverse industries and encounter a variety of uncertainties and load disturbances. Furthermore, model-free control strategies are particularly appealing for use in the mentioned actuation applications due to their ability to operate effectively without the need for precise system models. This characteristic simplifies setup and reduces computational demands, making these systems ideal for dynamic environments where system properties may change over time \cite{corradini2021robust, truong2022backstepping, hamon2023model, wang2023continuous}.
\\
\indent Hence, this paper aims to design a model-free, generic, robust control (GRC) framework for servo-driven actuators, ensuring strong stability and adaptability in the presence of uncertainties and load disturbances. To demonstrate the generality of the GRC framework: 1) we initially model the most prevalent types of the EDA, HDA, and PDA mechanisms and elucidate their generic dynamic forms. 2) Subsequently, the state-space model of these mechanisms is decomposed into smaller subsystems.
3) The GRC adopts subsystem-based control strategies, which facilitates robust control for each state-variant subsystem of the motion dynamics and the energy conversion in actuators separately, while adhering to control input constraints, regardless of the type of servo power. In summary, this paper provides the following contributions: 1) Addressing the energy conversion relationships, the GRC strategy is completely model-free and can be applied to any servo-driven actuation mechanism encompassing HDAs, PDAs, or EDAs. 2) The robustness and uniformly exponential stability in tracking desired motions are ensured. 3) All control signals generated by the GRC for each actuation subsystem are confined within predefined limitations. 4) The performance of the GRC is validated experimentally on two highly complex servo-driven actuators.

\section{Eda Mechanism}
An EDA system typically functions by utilizing an electric motor that converts electrical energy, such as a battery or power grid, into mechanical energy linked through a gearbox, facing nonlinearity and uncertainty \cite{jeong2022nonlinear}.
\subsection{Electric Motor Modeling}
\subsubsection{Universal Motor Modeling}
Universal motors are powered by a voltage, where the stator's magnetizing windings are in series with the rotor windings. The brushes ensure the rotor windings are mechanically fed. The dynamics of universal motors can be determined, as in \cite{abdeljawed2022simulation} and \cite{li2013adaptive}:
\begin{equation}
\small
\begin{aligned}
\label{eq1}
& \frac{d i}{dt}=\frac{1}{L_a+L_f} V - \frac{R_a+R_f}{L_a+L_f} i - \frac{1}{L_a+L_f} E\\
& \frac{d \omega}{d t}=\frac{1}{J_m} \phi_m i - \frac{1}{J_m} b_m \omega - \frac{1}{J_m} \tau_{L} - \frac{1}{J_m} \tau_{f s}
\end{aligned}
\end{equation}
where $i$ stands for the motor current, $V$ the supply voltage, and $\omega$ the angular shaft speed. The armature and field resistances are represented by $R_a$ and $R_f$, respectively, while $L_a$ and $L_f$ are the inductances of the armature and field, respectively. Further, $\tau_L$ is the torque of the load, $b_m$ is the coefficient of viscous friction, $J_m$ represents the motor's moment of inertia, and $\tau_{f s}$ is the friction torque. Both the counter electromotive force (EMF) $E$ and the electromagnetic torque generated $\tau_m$ are contingent on the flux $\phi_m$, as:

\begin{equation}
\small
\begin{aligned}
\label{eq2}
& E=\omega \phi_m, \hspace{0.2cm} \tau_m=i \phi_m
\end{aligned}
\end{equation}

\subsubsection{Three-Phase Motor Modeling}

The direct-quadrature (d-q) reference frame is a mathematical technique used to simplify the analysis and control of three-phase electric motors by converting the variable elements of a three-phase system into two orthogonal components \cite{chatri2022design}. Among these types of electric motors, permanent magnet synchronous motors (PMSMs) stand out for their higher power density, consistent torque output, and quieter operation \cite{rezaeizadeh2024reliability}.
Considering uncertainties in parameters, the mathematical representation of PMSMs' d-q frame is as follows \cite{wu2019robust}:
\begin{equation}
\small
\begin{aligned}
\label{eq3}
L_d \frac{d i_d}{d t}=&-R_s i_d+n_p \omega L_q i_q+n_p \omega \Delta L_q i_q+u_d-\Delta R_s i_d\\
&-\Delta L_d \dot{i}_d \\
L_q \frac{d i_q}{d t}=&-R_s i_q-n_p \omega L_d i_d-n_p \omega \phi_m+u_q-\Delta R_s i_q\\
&+n_p \omega \Delta L_d i_d-n_p \omega \Delta \phi_m-\Delta L_q \dot{i}_q 
\end{aligned}
\end{equation}
where $i_d$ and $i_q$ indicate the stator currents along the d- and q-axes, respectively, and $u_d$ and $u_q$ are the corresponding voltages on these axes. The inductances $L_d$ and $L_q$ occur along the d- and q-axes, with $R_s$ as the stator resistance. The number of pole pairs is denoted by $n_p$.
In addition, the errors between the theoretical and actual values are defined as the $\Delta R_s$, $\Delta L_d$, $\Delta L_q$, $\Delta J$, and $\Delta \phi_m$. The relationship between motor torque and currents is as follows \cite{jeong2022nonlinear, heydari2024robust}:
\begin{equation}
\small
\begin{aligned}
\label{eq4}
\tau_m=\frac{3}{2} n_p i_q\left(\phi_{m}+\left(L_d-L_q\right) i_d\right)
\end{aligned}
\end{equation}
From \cite{jeong2022nonlinear, heydari2024robust}, and \cite{chatri2022design}, the dynamical motion of the motor is:
\begin{equation}
\small
\begin{aligned}
\label{eq5}
J_m \frac{d \omega}{d t}&=\frac{3}{2} n_p i_q\left(\phi_{m}+\left(L_d-L_q\right) i_d\right)-\tau_{fs}-b_m \omega\\
&-\tau_L+n_p\left[\Delta \psi i_q+\left(\Delta L_d-\Delta L_q\right) i_d i_q\right]-\Delta J \dot{\omega}
\end{aligned}
\end{equation}

\subsection{Linear and Rotational Motor-Powered EDAs}
\begin{figure}[h!]
    \hspace*{-0.0cm} % Adjust the value as needed
    \centering
    \scalebox{0.95}{\includegraphics[trim={0cm 0.0cm 0.0cm 0cm},clip,width=\columnwidth]{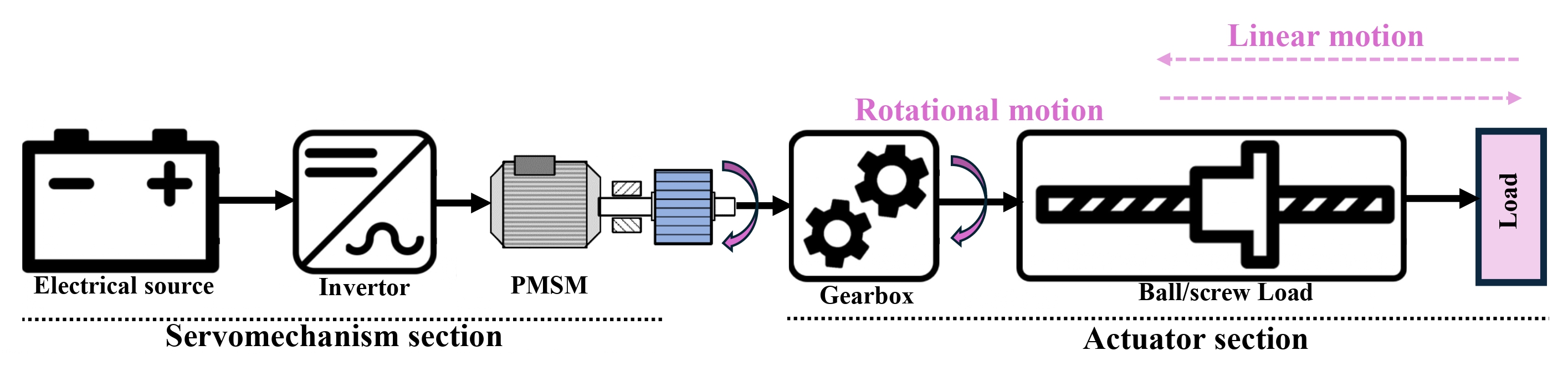}}
    \caption{The schematic of a linear EDA.}
    \label{fig2}
\end{figure}

For applications requiring rotational actuators, the gearbox output may directly drive the load side or may be further modified through additional gearing to adjust the output characteristics.
Alternatively, to convert rotational motion from a motor into precise linear motion in actuator mechanisms, a ball screw setup is commonly used. Fig. \ref{fig2} illustrates the schematic of an electromechanical linear actuator powered by a PMSM.

By considering the coefficient of converting rotary movement to linear movement, as well as the gear ratio, linear position of the actuator $x_L$, equivalent inertia $J_{e q}$, damping $b_{e q}$, spring effect $k_{e q}$, load coefficient $f_{e q}$ and load force $f_L$, we can adopt the mathematical formulation provided in \eqref{eq5} for linear motion dynamics \cite{heydari2024robust}, as:
\begin{equation}
\small
\begin{aligned}
\label{eq6}
\ddot{x}_L=\frac{1}{J_{e q}}\left[\frac{3}{2} n_p i_q\phi_{m}-b_{e q} \dot{x}_L-k_{e q} x_L-f_{e q} f_L\right]
\end{aligned}
\end{equation}
Similarly, we can produce the mathematical formulation for rotational motion dynamics, as:
\begin{equation}
\small
\begin{aligned}
\label{eq66}
\ddot{\theta}=\frac{1}{J_{e q}}\left[\frac{3}{2} n_p i_q\phi_{m}-b_{e q} \omega-k_{e q} \theta-f_{e q} f_L\right]
\end{aligned}
\end{equation}
where $\theta$ and $\omega$ are the angular position and velocity of the actuator, respectively.

\section{Hda Mechanism}
A hydraulic actuator system commonly operates by using a pump to move hydraulic fluid from a reservoir through hydraulic lines to a valve that controls the fluid's flow and direction. This pressurized fluid is then directed into an actuator, such as a cylinder or hydraulic motor, which converts the hydraulic energy into mechanical energy. After exerting force, the fluid returns to the reservoir, completing the cycle.
\subsection{Cylinder Modeling in HDAs}
\begin{figure}[h!]
    \hspace*{-0.0cm} % Adjust the value as needed
    \centering
    \scalebox{0.5}{\includegraphics[trim={0cm 0.0cm 0.0cm 0cm},clip,width=\columnwidth]{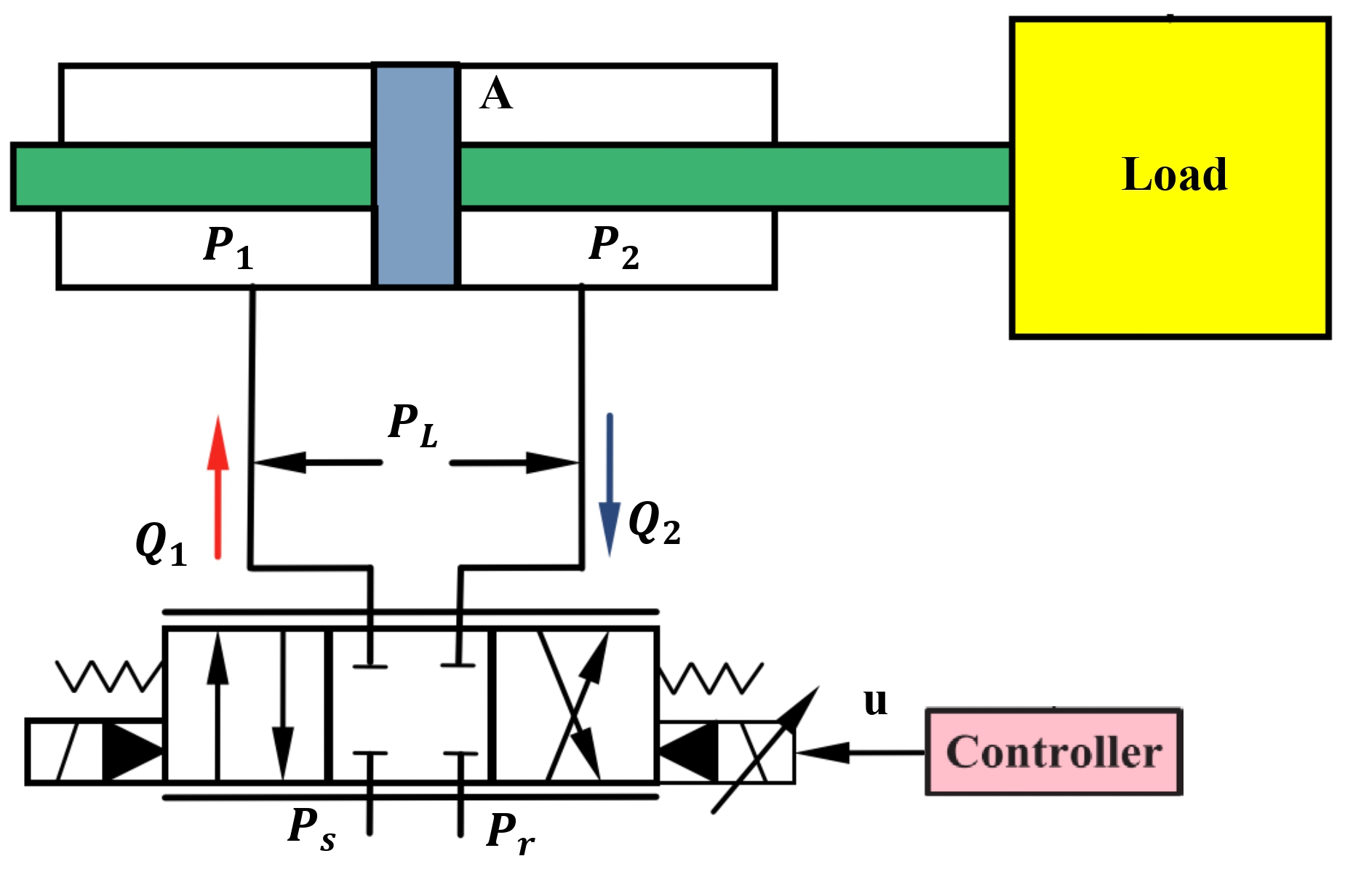}}
    \caption{The schematic of a linear HDA \cite{yao2017active}.}
    \label{fig3}
\end{figure}

Fig. \ref{fig3} illustrates a hydraulic servo system, featuring a double-rod hydraulic cylinder controlled by a servo valve, which drives an inertial load \cite{yao2017active}. $P_s$ refers to the input pressure of the fluid, while $P_r$ indicates the pressure at the return.

The flows entering and exiting the servo valve are $Q_1$ and $Q_2$. Regarding the oil's compressibility, the dynamics of the cylinder's pressure are described in \cite{yao2017active}:
\begin{equation}
\small
\begin{aligned}
\label{eq7}
& \frac{V_1}{\beta_e} \dot{P}_1=-A_h \dot{x}_L-C_t P_l+q_1(t)+Q_1 \\
& \frac{V_2}{\beta_e} \dot{P}_2=A_h \dot{x}_L+C_t P_l-q_2(t)-Q_2
\end{aligned}
\end{equation}
In this setup, the volumes of the cylinder chambers are given by $V_1$ and $V_2$, and $x_L$ is the displacement of the load. The term $\beta_e$ refers to the effective oil bulk modulus, and $C_t$ denotes the coefficient of the total internal leakage of the cylinder, influenced by pressure differences $P_l = P_1 - P_2$, where $P_1$ and $P_2$ are the respective chamber pressures. The terms $q_1(t)$ and $q_2(t)$ are the modeling inaccuracies, where $Q_1$ and $Q_2$ can be defined, as \cite{manring2019hydraulic}:

\begin{equation}
\small
\begin{aligned}
\label{eq8}
& Q_1=k_u u\left[s(u) \sqrt{P_s-P_1}+s(-u) \sqrt{P_1-P_r}\right] \\
& Q_2=k_u u\left[s(u) \sqrt{P_2-P_r}+s(-u) \sqrt{P_s-P_2}\right]
\end{aligned}
\end{equation}
where $k_u$ represents the total flow gain relative to the control input $u$, and the function $s(u)$ is characterized as:
\begin{equation}
\small
\begin{aligned}
\label{eq9}
s(u)=\left\{\begin{array}{l}
1, \text { if } u \geq 0 \\
0, \text { if } u<0
\end{array}\right.
\end{aligned}
\end{equation}
In applications with a high-response servo valve, the dynamics of the valve can often be overlooked \cite{yao2017active}. In such an instance, the equation representing the force equilibrium of the inertial load can be expressed as follows:
\begin{equation}
\small
\begin{aligned}
\label{eq10}
J_{h} \ddot{x}_L=P_l D_h-b_h \dot{x}_L-A_f S_f(\dot{x}_L)+D_L
\end{aligned}
\end{equation}
where $J_{h}$ denotes the mass of the load. The effective area of the cylinder's ram is represented by $D_h$, while the parameter $b_h$ captures the combined effects of modeled damping and viscous friction acting on both the load and cylinder rod. The term $A_f S_f$ approximates nonlinear Coulomb friction, with $A_f$ being the amplitude of the Coulomb friction and $S_f$ representing a predetermined shape function \cite{yao2017active}. Lastly, $D_L$ accounts for additional disturbances.

\subsection{Hydraulic Motor Modeling}
\begin{figure}[h!]
    \hspace*{-0.0cm} % Adjust the value as needed
    \centering
    \scalebox{0.65}{\includegraphics[trim={0cm 0.0cm 0.0cm 0cm},clip,width=\columnwidth]{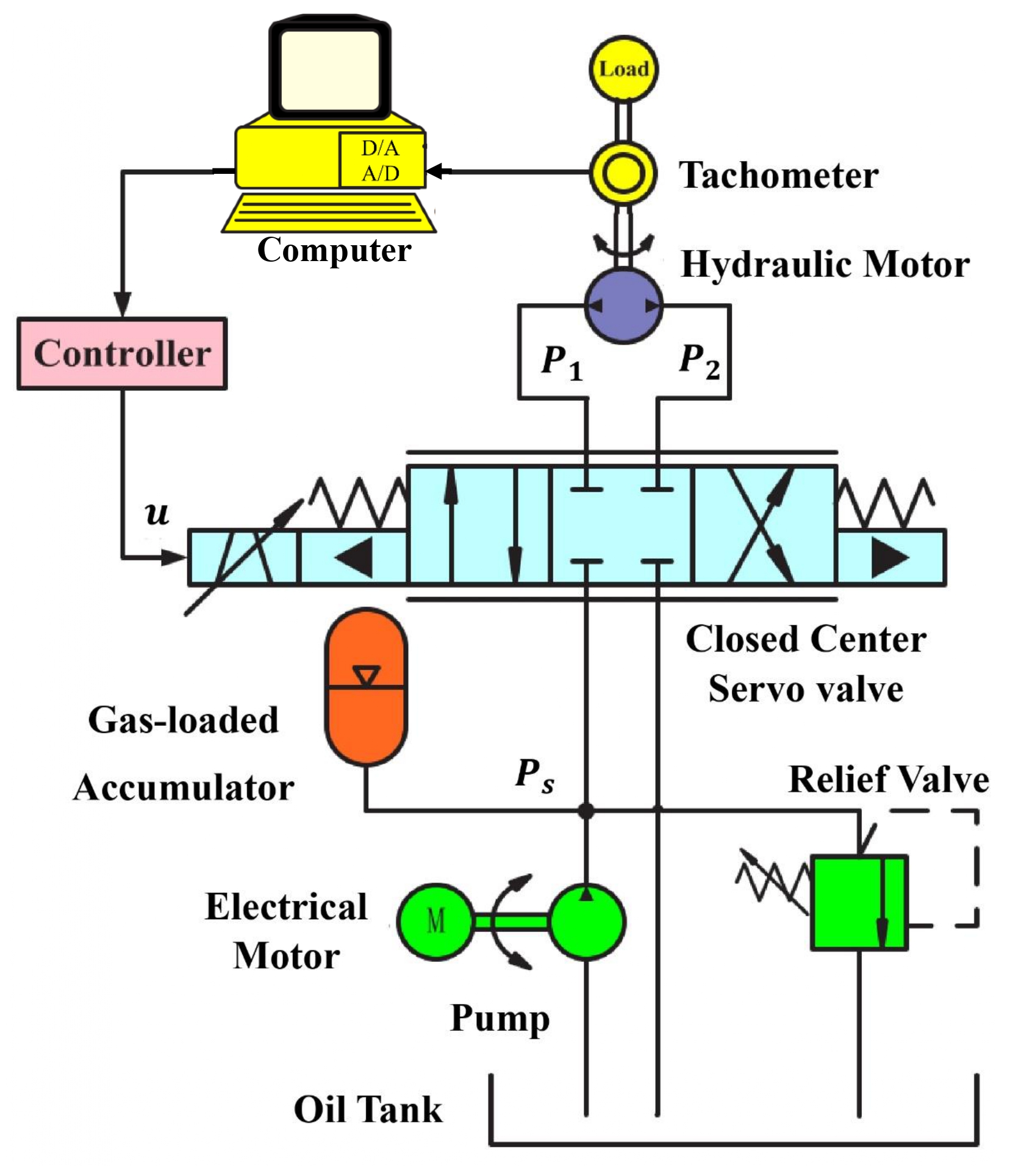}}
    \caption{The schematic of a rotational HDA \cite{yang2017position}.}
    \label{fig4}
\end{figure}

Fig. \ref{fig4} presents a model of an electro-hydraulic servo system via a torque meter, tachometer, and pressure sensors.
For a thorough investigation and generality, in this section, we do not neglect the valve dynamic, and the movement of a servo valve spool can be defined, as \cite{yang2017position}:
\begin{equation}
\small
\begin{aligned}
\label{eq11}
\tau_v W \dot{x}_v+W x_v=K_v u
\end{aligned}
\end{equation}
where $\tau_v$ is a constant, $W$ indicates the area gradient of the servo valve, $x_v$ refers to the spool displacement in the servo valve, $K_v$ symbolizes the constant gain of the servo valve, and $u$ is the control input.
The flow rate entering and exiting the servo valve, denoted as $Q = Q_1 - Q_2$, can be \cite{yang2017position}:
\begin{equation}
\small
\begin{aligned}
\label{eq12}
Q=C_d W x_v \sqrt{\frac{P_s-\operatorname{sign}\left(W x_v\right) P_l}{\rho}}
\end{aligned}
\end{equation}
where $C_d$ is identified as the flow discharge coefficient and $\rho$ is the mass density of the fluid oil. We can denote the relationship between $P_l$ and the servo valve spool, as \cite{yang2017position}: 
\begin{equation}
\small
\begin{aligned}
\label{eq13}
\frac{V}{2 \beta_e} \dot{P}_l=C_d W x_v \sqrt{\frac{P_s-\operatorname{sign}\left(W x_v\right) P_l}{\rho}}-D_{eh} \dot{\theta}-C_t P_l
\end{aligned}
\end{equation}
where $V$ refers to the volume of oil in one of the actuator's chambers, $D_{eh}$ is the volumetric displacement of the actuator, and $\theta$ is the output angular position. The equation for motion can be expressed as follows \cite{yang2017position}:
\begin{equation}
\small
\begin{aligned}
\label{eq14}
J_{eh} \ddot{\theta}=P_l D_{eh}-b_{eh} \dot{\theta}+D_L
\end{aligned}
\end{equation}
$J_{eh}$ is the inertia of the actuator, $b_{eh}$ is the coefficient of viscous damping, and $D_L$ indicates the load torque.\\
\indent \textbf{Remark 1:} In Section 3-A, as in \cite{yao2017active}, we neglected the spool dynamics due to the rapid response characteristics of a high-response servo valve, which operates faster than other system components. However, in Section 3-B, as in \cite{yang2017position}, we considered the spool dynamics to be a separate subsystem. We aim to provide a generic control solution for actuators; therefore, we investigated both scenarios separately.

\section{PDA Mechanism}
\begin{figure}[h!]
    \hspace*{-0.0cm} % Adjust the value as needed
    \centering
    \scalebox{0.7}{\includegraphics[trim={0cm 0.0cm 0.0cm 0cm},clip,width=\columnwidth]{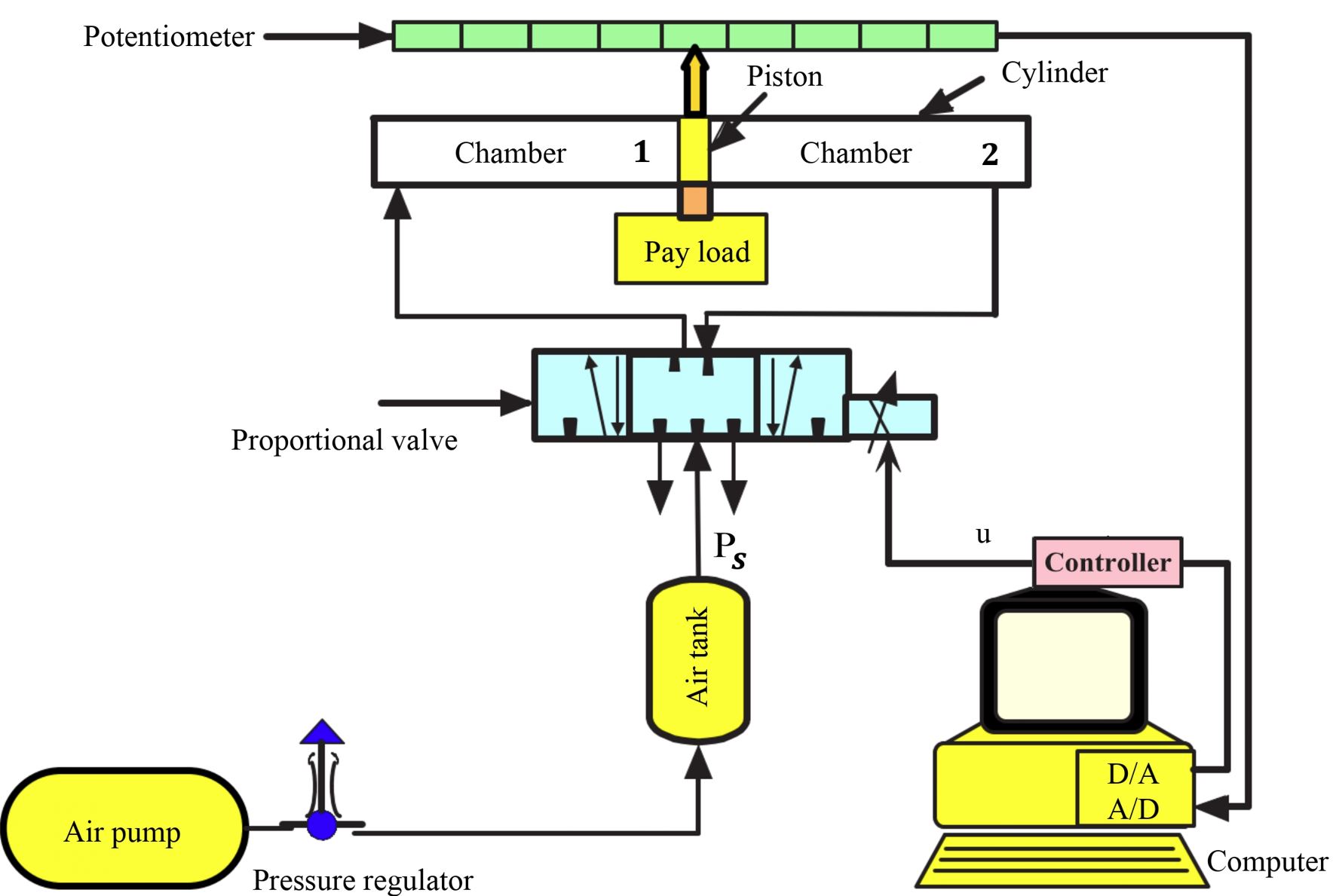}}
    \caption{The schematic of a linear PDA \cite{ren2019adaptive}.}
    \label{fig5}
\end{figure}

This mechanism closely resembles HDAs, albeit with different types of energy and power. For instance, Fig. \ref{fig5} illustrates how a pneumatic servo system operates \cite{ren2019adaptive}. Compressed air is supplied by an air pump, and the flow into Chambers 1 and 2 is managed by a proportional valve. This valve adjusts the air pressure in both chambers based on control signals, facilitating the movement of the payload, monitored by a potentiometer \cite{wang2023position}.
The representation of the pneumatic servo system can be described as \cite{ren2019adaptive}:
\begin{equation}
\small
\begin{aligned}
\label{eq15}
\begin{array}{l}
\dot{m}_1=g_1\left(u, P_1\right), \hspace{0.2cm}\dot{m}_2=g_2\left(u, P_2\right) \\
K_p R_p T_p \dot{m}_1=K_p P_1 A_1 \dot{x}_L+A_1\left(x_0+x_L\right) \dot{P}_1 \\
K_p R_p T_p \dot{m}_2=-K_p P_2 A_2 \dot{x}_L+A_2\left(x_0-x_L\right) \dot{P}_2 \\
J_p \ddot{x}_L=P_1 A_1-P_2 A_2-F_f
\end{array}
\end{aligned}
\end{equation}

\hspace{-0.35cm}where $\dot{m}_1$ and $\dot{m}_2$ denote the mass flow rates of gas into Chambers 1 and 2, respectively. The piston areas facing Chambers 1 and 2 are $A_1$ and $A_2$, respectively, and $x_L$ represents the displacement of the payload, while $x_0$ indicates the initial position. The total mass of the payload and piston is given by $J_p$, and $F_f$ refers to the friction force encountered. The specific heat ratio is noted as $K_p$, with $R_p$ symbolizing the ideal gas constant and $T_p$ the temperature of the air. The input voltage to the proportional valve is $u$. The functions $g_1(u, P_1)$ and $g_2(u, P_2)$ are nonlinear expressions that depend on the pressures upstream and downstream in Chambers 1 and 2, respectively, and are specified as follows \cite{ren2019adaptive}:

\begin{equation}
\small
\begin{aligned}
\label{eq16}
\begin{array}{l}
g_1\left(u, P_1\right)=\sqrt{P_s-P_1}\left(c_{11} u+c_{12} u^2\right) \\
g_2\left(u, P_2\right)=\sqrt{P_2-P_0}\left(c_{21} u+c_{22} u^2\right)
\end{array}
\end{aligned}
\end{equation}
where $p_0$ denotes the atmospheric pressure, and $c_{11}, c_{12}, c_{21}$, and $c_{22}$ are constants associated with the properties of air. To aid in the analysis and system design, the functions $g_1$ and $g_2$ can be linearized along with the friction $F_f$ \cite{ren2019adaptive}. This process simplifies the nonlinear mathematical model into a third-order linear model, as follows \cite{ren2019adaptive}:
\begin{equation}
\small
\label{eq17}
\begin{array}{l}
\dddot{x}_L=a_1 x_L+a_2 \dot{x}_L+a_3 \ddot{x}_L +bu+b\Delta u+d
\end{array}
\end{equation}
where, $a_1, a_2$, and $a_3$ represent unknown parameters; $b$ denotes the unknown control gain; $d$ indicates the disturbance; and $\Delta u$ refers to the neutral position of the proportional valve.

\section{Unifying Formulation and Control Design}
\subsection{Generic Servo-Driven Actuation System Model}
Before designing the proposed model-free control framework, the dynamic equations provided for all mentioned actuators in Eqs. \eqref{eq6}, \eqref{eq66}, \eqref{eq10}, \eqref{eq14}, and \eqref{eq17}, as well as the energy conversion equations described in Eqs. \eqref{eq1}, \eqref{eq3}, \eqref{eq7}, \eqref{eq11}, \eqref{eq13}, and \eqref{eq15}, can be unified as follows:

\begin{equation}
\small
\label{eq18}
\begin{array}{l}
\dot{x}_1(t)=\alpha_1 x_2(t) + F_1 (x,t) + D_1 (t) \\
\dot{x}_2(t)=\alpha_2 Sat_1(u_1(t)) + F_2 (x,t) + D_2 (t)  \\
\dot{x}_3(t)=\alpha_3 Sat_2(u_2(t)) + F_3 (x,t) + D_3 (t)  \\
\dot{x}_4(t) = \alpha_4 Sat_3(u_3(t)) + F_4 (x,t) + D_4 (t) 
\end{array}
\end{equation}
where $\alpha_{1}$,...,$\alpha_4$ are unknown positive coefficients. $F_1$,..., $F_4$ are unknown state-variant modeling and uncertainty terms. $D_1$,...,$D_4$ are time-variant external disturbances and load effects. By defining $\upsilon = 0, 1, 2, 3$, $Sat_\upsilon(u_\upsilon) = s_{1\upsilon} u_\upsilon(t)+s_{2\upsilon}$ is a constraining function of the control input $u_\upsilon$, which was defined as \cite{shahna2023exponential}:

\begin{equation}
\small
\begin{aligned}
\label{eq19}
&\begin{aligned}
& s_{1\upsilon}= \begin{cases}\frac{1}{|u_\upsilon(t)|+1}, & \text { if } u_\upsilon(t) \geq u_{max} \text { or } u_\upsilon(t) \leq u_{min} \\
1 & \text { if } u_{min} \leq u_\upsilon(t) \leq u_{max}\end{cases} \\
& s_{2\upsilon}= \begin{cases}u_{min}-\frac{u_\upsilon(t)}{|u_\upsilon(t)|+1}, & \text { if } u_\upsilon(t) \geq u_{min} \\
0 & \text { if } u_{min} \leq u_\upsilon(t) \leq u_{max} \\
u_{max}-\frac{u_\upsilon(t)}{|u_\upsilon(t)|+1} & \text { if } u_\upsilon(t) \leq u_{max}\end{cases}
\end{aligned}
\end{aligned}
\end{equation}
$u_0$ is a virtual control. Note that $s_{10}=1$, $s_{1\upsilon} > 0$ and $s_{20}=0$. To clarify, the decomposed model is shown in Table \ref{table1}.

\begin{table}[h!]
  \caption{Servo-driven actuator modeling parameters}
  \centering
  \scriptsize
  \begin{tabular}{cccc}
    \toprule
    \toprule
    \textcolor{black}{\textbf{Servo system}} &\textcolor{black}{\textbf{Defined}} & 
    \textcolor{black}{\textbf{Reference}} &
    \textcolor{black}{\textbf{Control}}  \\
        \textcolor{black}{\textbf{types}} &\textcolor{black}{\textbf{states}} &
    \textcolor{black}{\textbf{trajectory}} &
    \textcolor{black}{\textbf{input}} \\
    \midrule
                           \multirow{2}{*}{\textbf{\textcolor{black}{Single-phase motor}}} &\textcolor{black}{{$x_1 = \theta/x_L$}} & \textcolor{black}{$x_{1d} = x_d$} &   $u_0$ \\
    \multirow{2}{*}{\textbf{\textcolor{black}{EDAs}}} & \textcolor{black}{$x_2 = \omega/\dot{x}_L$}  & \textcolor{black}{$x_{2d} = \dot{x}_d$} & $u_1$  \\
       & \textcolor{black}{$x_3 = i$} & \textcolor{black}{$x_{3d} = u_1$} & $u_2 = V$  \\
                      \hdashline
                           \multirow{3}{*}{\textbf{\textcolor{black}{Three-phase motor}}} &\textcolor{black}{{$x_1 = \theta/x_L$}} & \textcolor{black}{$x_{1d} = x_d$} & \textcolor{black}{$u_0$}  \\
    & \textcolor{black}{$x_2 = \omega/\dot{x}_L$} & \textcolor{black}{$x_{2d} = \dot{x}_d$} & \textcolor{black}{$u_1$}  \\
 \multirow{1}{*}{\textbf{\textcolor{black}{EDAs}}} & \textcolor{black}{$x_3 = i_q$} & \textcolor{black}{$x_{3d} = u_1$} & \textcolor{black}{$u_2 = u_q$}  \\
                   & \textcolor{black}{$x_4 = i_d$}  & \textcolor{black}{$x_{4d} = 0$} & \textcolor{black}{$u_3 = u_d$} \\
                   \midrule
                           \multirow{2}{*}{\textbf{\textcolor{black}{HDA}}} &\textcolor{black}{{$x_1 = \theta/x_L$}}  & \textcolor{black}{$x_{1d} = x_d$} & \textcolor{black}{$u_0$}  \\
    \multirow{2}{*}{\textbf{\textcolor{black}{without valve dynamics}}} & \textcolor{black}{$x_2 = \omega/\dot{x}_L$} & \textcolor{black}{$x_{2d} = \dot{x}_d$} & \textcolor{black}{$u_1$}  \\
       & \textcolor{black}{$x_3 = P_l$} & \textcolor{black}{$x_{3d} = u_1$} & \textcolor{black}{$u_2 = u$}  \\
                      \hdashline
                           \multirow{3}{*}{\textbf{\textcolor{black}{HDA}}} &\textcolor{black}{{$x_1 = \theta/x_L$}} & \textcolor{black}{$x_{1d} = x_d$} & \textcolor{black}{$u_0$}  \\
    &\textcolor{black}{{$x_2 = \omega/\dot{x}_L$}} & \textcolor{black}{$x_{2d} = \dot{x}_d$} & \textcolor{black}{$u_1$}  \\
       \multirow{1}{*}{\textbf{\textcolor{black}{with valve dynamics}}}&\textcolor{black}{{$x_3 = P_l$}}  & \textcolor{black}{$x_{3d} = u_1$} & \textcolor{black}{$u_2$}  \\
                   &\textcolor{black}{{$x_4 = W x_v$}} & \textcolor{black}{$x_{4d} = u_2$} & \textcolor{black}{$u_3 = u$} \\
                       \midrule
                           \multirow{2}{*}{\textbf{\textcolor{black}{PDA}}} &\textcolor{black}{{$x_1 = \theta/x_L$}} &  \textcolor{black}{$x_{1d} = x_d$} & \textcolor{black}{$u_0$}  \\
    \multirow{2}{*}{\textbf{\textcolor{black}{without valve dynamics}}} & \textcolor{black}{$x_2 = \theta/\dot{x}_L$} & \textcolor{black}{$x_{2d} = \dot{x}_d$} & \textcolor{black}{$u_1$}  \\
       & \textcolor{black}{$x_3 = \theta/\ddot{x}_L$} & \textcolor{black}{$x_{3d} = \ddot{x}_d$} & \textcolor{black}{$u_2 = u$}  \\
                      \hdashline
                           \multirow{3}{*}{\textbf{\textcolor{black}{PDA}}} &\textcolor{black}{{$x_1 = \theta/x_L$}} &  \textcolor{black}{$x_{1d} = x_d$} & \textcolor{black}{$u_0$}  \\
    &\textcolor{black}{{$x_2 = \dot{\theta}/\dot{x}_L$}} & \textcolor{black}{$x_{2d} = \dot{x}_d$} & \textcolor{black}{$u_1$}  \\
       \multirow{1}{*}{\textbf{\textcolor{black}{with valve dynamics}}}&\textcolor{black}{{$x_3 = u_1 = \ddot{\theta}/\ddot{x}_L$}}  & \textcolor{black}{$x_{3d} = \ddot{x}_d$} & \textcolor{black}{$u_2$}  \\
                   &\textcolor{black}{{$x_4 = W x_v$}} & \textcolor{black}{$x_{4d} = u_2$} & \textcolor{black}{$u_3 = u$} \\
    \bottomrule
    \bottomrule
  \end{tabular}
  
  \label{table1}
  \begin{tablenotes} 
\item[-]- Depending on whether the actuator motion is rotational or linear, the position and velocity states ( $x_1$ and $x_2$ ) are selected as either angular position and velocity $(\theta$ and $\dot{\theta} = \omega$ ), or linear position and velocity $\left(x_L\right.$ and $\left.\dot{x}_L\right)$.
\item[-]- The energy conversion states that drive the actuators in EDAs are based on currents, while in HDAs and PDAs, they are based on oil and air pressures.
\item[-]- Single-phase motor EDAs, as well as both HDAs and PDAs with negligible valve dynamics, consist of three subsystems and do not require a fourth.
\item[-]- According to the Park transformation commonly referenced in the control literature, controlling the system is easier when the d-axis current state ($i_d$) tracks zero. Thus, $x_{4d} = i_{d(\text{ref})} = 0$ in three-phase motor EHAs \cite{heydari2024robust}.
\end{tablenotes}
\end{table}

\subsection{The GRC Framework}
Suppose we define $j$ as the number of subsystems, ranging from $j=1$ to $j=3$ for single-phase motor-powered EDAs, as well as for HDAs and PDAs with negligible valve dynamics, or from $j=1$ to $j=4$ for the remaining actuation mechanisms mentioned. In this case, the tracking error for the subsystem-based model of actuator mechanisms can be defined as follows:

\begin{equation}
\small
\begin{aligned}
\label{eq20}
& e_j (t) = x_j - x_{jd}, \hspace{0.2cm} z_j= \begin{cases}e_j -u_0, & \text { if } j=2 \\
e_j &  \text{else} \end{cases} 
\end{aligned}
\end{equation}
where $x_{jd}$ is the desired signal for state $x_j$. Meanwhile, $z_j$ is the tracking transform, and $u_0$ is the virtual control. The subsystem-based model-free GRC can be defined for all subsystems as:
\begin{equation}
\small
\begin{aligned}
\label{eq21}
u_{\upsilon}= \begin{cases}-\frac{1}{2}\left(k_{\upsilon}+\epsilon_{\upsilon} \hat{\chi}_{\upsilon}\right) z_{\upsilon+1} - z_{\upsilon}, & \text { if } \upsilon=1 \\
-\frac{1}{2}\left(k_{\upsilon}+\epsilon_{\upsilon} \hat{\chi}_{\upsilon}\right) z_{\upsilon+1} & \text{else} \end{cases} 
\end{aligned}
\end{equation}
where $k_\upsilon$, and $\epsilon_\upsilon$ are positive constants. $\hat{\chi}_\upsilon$ is an adaptive law parameter used to adjust the control $u_\upsilon$ for generating a sufficient signal to track the reference trajectory in each subsystem of the servo-driven actuators, even under uncertainty and disturbances in model-free mechanisms. The subsystem-based adaptive laws can be proposed as follows:
\begin{equation}
\small
\begin{aligned}
\label{eq22}
&\frac{d \hat{\chi}_\upsilon}{d t}=-\gamma_\upsilon \delta_\upsilon \hat{\chi}_\upsilon+\frac{1}{2} \epsilon_\upsilon \gamma_\upsilon\left|z_{\upsilon+1}\right|^2
\end{aligned}
\end{equation}
where $\gamma_\upsilon$, and $\delta_\upsilon$ are positive constants.
By considering Eqs. \eqref{eq18}, \eqref{eq19}, and \eqref{eq20}, we can produce a state-space model of the tracking transformation, as:
\begin{equation}
\small
\begin{aligned}
\label{eq23}
\dot{z}_1 =& \alpha_1 z_2 + \alpha_1 u_0 + F_1 + D^*_1\\
\dot{z}_2 =& \alpha_2 s_{11} u_1 + F^*_2 + D^*_2\\
\dot{z}_3 =& \alpha_3 s_{12} u_2 + F_3 + D^*_3\\
\dot{z}_4 =& \alpha_4 s_{13} u_3 + F_4 + D^*_4
\end{aligned}
\end{equation}
where $D^*_1 = D_1 - (1+\alpha_1) \dot{x}_{1d}$, $F^*_2 = F_2 - \dot{u}_0 $, $D^*_2 = D_2 - \dot{x}_{2d} + \alpha_2 s_{21}$, $D^*_3 = D_3 - \dot{x}_{3d} + \alpha_3 s_{22}$, and $D^*_4 = D_4 - \dot{x}_{4d} + \alpha_4 s_{23}$. The adaptive error is defined as $\tilde{\chi}_\upsilon = \hat{\chi}_\upsilon - \chi^*_\upsilon$ where $\chi^*_\upsilon$ is an unknown positive constant. 
By defining adaptive law error as $\tilde{\chi}_\upsilon=\hat{\chi}_\upsilon-\chi^*_\upsilon$, and from $\dot{\hat{\chi}}_\upsilon$, we can obtain:
\begin{equation}
\small
\begin{aligned}
\label{eq22}
&\frac{d \tilde{\chi}_\upsilon}{d t}=-\gamma_\upsilon \delta_\upsilon \tilde{\chi}_\upsilon+\frac{1}{2} \epsilon_\upsilon \gamma_\upsilon\left|z_{\upsilon+1}\right|^2 - \gamma_\upsilon \delta_\upsilon {\chi}^*_\upsilon
\end{aligned}
\end{equation}
Now, we introduce a Lyapunov function for the first subsystem of the actuation system, as follows:
\begin{equation}
\small
\begin{aligned}
\label{eq24}
{V}_{1} =\frac{1}{2} \hspace{0.1cm} [\alpha_{1}^{-1} s_{10} ^{-1}{z^2_{1}}+{\gamma^{-1}_{0}}\tilde{\chi}_{0}^2]
\end{aligned}
\end{equation} 
Note that in the first subsystem, $s_{10}=1$. Thus, after the derivative, we have:
\begin{equation}
\small
\begin{aligned}
\label{eq25}
\dot{V}_1 = \alpha_{1}^{-1}\dot{z}_1 {z_{1}}+{\gamma^{-1}_{0}}\dot{\tilde{\chi}}_0\tilde{\chi}_{0}
\end{aligned}
\end{equation} 
From the $\dot{z}_1$ definition, we have:
\begin{equation}
\small
\begin{aligned}
\label{eq25}
\dot{V}_1 =& {z_{1}} z_2  + {z_{1}} u_0 + \alpha_{1}^{-1} F_1 {z_{1}} + \alpha_{1}^{-1} D^*_1 {z_{1}} - \delta_0 \tilde{\chi}^2_0\\
&+\frac{1}{2} \epsilon_0 \left|z_{1}\right|^2 \tilde{\chi}_{0}
-  \delta_0 {\chi}^*_0 \tilde{\chi}_{0}
\end{aligned}
\end{equation} 
Assume $\zeta_0$ and ${d}_{max1} \in \mathbb{R}^+$ are unknown positive constants, and $r_0: \mathbb{R} \rightarrow \mathbb{R}^+$ is a continuously bounded function with strictly positive values. Then, according to the assumption that uncertainties and disturbances are bounded, we can have:
\begin{equation}
\small
\begin{aligned}
\label{eq35}
&\|\alpha_{1}^{-1} F_1\| \hspace{0.1cm} \leq \zeta_0 \hspace{0.1cm}r_0\hspace{0.1cm}, \hspace{0.2cm} \|\alpha_{1}^{-1} {D_1}^*\| \hspace{0.1cm} \leq {d}_{max1}
\end{aligned}
\end{equation}
From (28) and (29), we have:
\begin{equation}
\small
\begin{aligned}
\label{eq25}
\dot{V}_1 \leq & {z_{1}} z_2  + {z_{1}} u_0 + \zeta_0 r_0 |{z_{1}}| + {d}_{max1} |{z_{1}}| - \delta_0 \tilde{\chi}^2_0 \\
&+\frac{1}{2} \epsilon_0 \left|z_{1}\right|^2 \tilde{\chi}_{0}
-  \delta_0 {\chi}^*_0 \tilde{\chi}_{0}
\end{aligned}
\end{equation} 
By substituting $u_0$, we obtain:
\begin{equation}
\small
\begin{aligned}
\label{eq25}
\dot{V}_1 \leq& {z_{1}} z_2  -\frac{1}{2} k_{0} z^2_{1}-\frac{1}{2}\epsilon_{0} \hat{\chi}_{0} z^2_{1} + \zeta_0 r_0 |{z_{1}}| + {d}_{max1} |{z_{1}}| \\
&- \delta_0 \tilde{\chi}^2_0+\frac{1}{2} \epsilon_0 \left|z_{1}\right|^2 \tilde{\chi}_{0}
+  \delta_0 {\chi}^*_0 \tilde{\chi}_{0}
\end{aligned}
\end{equation} 
As with $\tilde{\chi}=\hat{\chi}-\chi^*$:
\begin{equation}
\small
\begin{aligned}
\label{eq25}
\dot{V}_1 \leq& {z_{1}} z_2  -\frac{1}{2} k_{0} z^2_{1}-\frac{1}{2}\epsilon_{0} {\chi}^*_{0} z^2_{1} + \zeta_0 r_0 |{z_{1}}| + {d}_{max1} |{z_{1}}| \\
&- \delta_0 \tilde{\chi}^2_0
-  \delta_0 {\chi}^*_0 \tilde{\chi}_{0}
\end{aligned}
\end{equation} 
Assume $\mu_0$ and $\lambda_0$ are any positive constants. By mathematical manipulations and Young's inequality, we obtain:
\begin{equation}
\small
\begin{aligned}
\label{eq25}
\dot{V}_1 \leq& {z_{1}} z_2  -\frac{1}{2} k_{0} z^2_{1}-\frac{1}{2}\epsilon_{0} {\chi}^*_{0} z^2_{1} + \frac{1}{4}\mu^{-1}_0 r^2_0 + \mu_0 \zeta^2_0 {z_{1}}^2   \\
&+\frac{1}{4} \lambda_0^{-1} + \lambda_0 {d}^2_{max1} {z_{1}}^2- \delta_0 \tilde{\chi}^2_0
-  \delta_0 {\chi}^*_0 \tilde{\chi}_{0}
\end{aligned}
\end{equation} 
Now, we can introduce a positive parameter that the adaptive is supposed to estimate, as:
\begin{equation}
\small
\begin{aligned}
\label{eq25}
\chi_0^* = \frac{2}{\epsilon_0}(\mu_0 \zeta^2_0+ \lambda_0 {d}^2_{max1})
\end{aligned}
\end{equation} 
From (33) and (34):
\begin{equation}
\small
\begin{aligned}
\label{eq25}
\dot{V}_1 \leq& {z_{1}} z_2  -\frac{1}{2} k_{0} z^2_{1} + \frac{1}{4}\mu^{-1}_0 r^2_0 +\frac{1}{4} \lambda_0^{-1} - \delta_0 \tilde{\chi}^2_0\\
&-  \delta_0 {\chi}^*_0 \tilde{\chi}_{0}
\end{aligned}
\end{equation} 
Then:
\begin{equation}
\small
\begin{aligned}
\label{eq25}
\dot{V}_1 \leq& {z_{1}} z_2  -\frac{1}{2} k_{0} z^2_{1} + \frac{1}{4}\mu^{-1}_0 r^2_0 +\frac{1}{4} \lambda_0^{-1} -\frac{1}{2} \delta_0 \tilde{\chi}^2_0\\
&-\frac{1}{2} \delta_0 \tilde{\chi}^2_0 -  \delta_0 {\chi}^*_0 \tilde{\chi}_{0}
\end{aligned}
\end{equation} 
Because $\tilde{\chi}=\hat{\chi}-\chi^*$:
\begin{equation}
\small
\begin{aligned}
\label{eq25}
\dot{V}_1 \leq& {z_{1}} z_2  -\frac{1}{2} k_{0} z^2_{1} + \frac{1}{4}\mu^{-1}_0 r_0 +\frac{1}{4} \lambda_0^{-1} -\frac{1}{2} \delta_0 \tilde{\chi}^2_0\\
&-\frac{1}{2} \delta_0 (\hat{\chi}_{0}-{\chi}^*_{0})^2 -  \delta_0 {\chi}^*_0 (\hat{\chi}_{0}-{\chi}^*_{0})
\end{aligned}
\end{equation} 
After simplifying, we have:
\begin{equation}
\small
\begin{aligned}
\label{eq25}
\dot{V}_1 \leq& {z_{1}} z_2  -\frac{1}{2} k_{0} z^2_{1} + \frac{1}{4}\mu^{-1}_0 r_0 +\frac{1}{4} \lambda_0^{-1} -\frac{1}{2} \delta_0 \tilde{\chi}^2_0\\
&+\frac{1}{2} \delta_0 {{\chi}^*_{0}}^2
\end{aligned}
\end{equation} 
From (26), we have:
\begin{equation}
\small
\begin{aligned}
\label{eq25}
\dot{V}_1 \leq& {z_{1}} z_2  - \Psi_1 V_1 + \frac{1}{4}\mu^{-1}_0 r_0 + \tilde{\mu}_1
\end{aligned}
\end{equation} 
where:
\begin{equation}
\small
\begin{aligned}
\label{eq25}
\Psi_1 = \min [k_0, \hspace{0.2cm} \delta_0 \gamma_0], \hspace{0.2cm} \tilde{\mu}_1 = \frac{1}{4} \lambda_0^{-1}+\frac{1}{2} \delta_0 {{\chi}^*_{0}}^2
\end{aligned}
\end{equation} 
Now, similar to (26), we can define another Lyapunov function for the second subsystem as:
\begin{equation}
\small
\begin{aligned}
\label{eq24}
{V}_{2} =\frac{1}{2} \hspace{0.1cm} [\alpha_{2}^{-1} s_{11} ^{-1}{z^2_{2}}+{\gamma^{-1}_{1}}\tilde{\chi}_{1}^2]
\end{aligned}
\end{equation} 
After the derivative:
\begin{equation}
\small
\begin{aligned}
\label{eq25}
\dot{V}_2 = \alpha_{2}^{-1} s_{11} ^{-1} \dot{z}_2 {z_{2}}+{\gamma^{-1}_{1}}\dot{\tilde{\chi}}_1\tilde{\chi}_{1}
\end{aligned}
\end{equation} 
By inserting $\dot{z}_2$, we obtain:
\begin{equation}
\small
\begin{aligned}
\label{eq25}
\dot{V}_2 =& {z_{2}} u_1 + \alpha_{2}^{-1} s_{11} ^{-1} F^*_2 {z_{2}} + \alpha_{2}^{-1} s_{11} ^{-1} D^*_2 {z_{2}} \\
&- \delta_1 \tilde{\chi}^2_1+\frac{1}{2} \epsilon_1 \left|z_{2}\right|^2 \tilde{\chi}_{1}
-  \delta_1 {\chi}^*_1 \tilde{\chi}_{1}
\end{aligned}
\end{equation} 
Assume $\zeta_1$ and ${d}_{max2} \in \mathbb{R}^+$ are unknown positive parameters, and $r_1: \mathbb{R} \rightarrow \mathbb{R}^+$ is a continuously bounded function with strictly positive values. According to  the assumption that uncertainties and disturbances are bounded, we can have:
\begin{equation}
\small
\begin{aligned}
\label{eq35}
&\|\alpha_{2}^{-1} s_{11} ^{-1} F^*_2\| \hspace{0.1cm} \leq \zeta_1 \hspace{0.1cm}r_1\hspace{0.1cm}, \hspace{0.2cm} \|\alpha_{2}^{-1} s_{11} ^{-1} {D_2}^*\| \hspace{0.1cm} \leq {d}_{max2}
\end{aligned}
\end{equation}
From (43) and (44), we obtain:
\begin{equation}
\small
\begin{aligned}
\label{eq25}
\dot{V}_2 \leq & {z_{2}} u_1 + \zeta_1 r_1 |{z_{2}}| + {d}_{max2} |{z_{2}}| - \delta_1 \tilde{\chi}^2_1+\frac{1}{2} \epsilon_1 \left|z_{2}\right|^2 \tilde{\chi}_{1}\\
&
-  \delta_1 {\chi}^*_1 \tilde{\chi}_{1}
\end{aligned}
\end{equation} 
By inserting $u_1$:
\begin{equation}
\small
\begin{aligned}
\label{eq25}
\dot{V}_2 \leq& -{z_{1}} z_2  -\frac{1}{2} k_{1} z^2_{2}-\frac{1}{2}\epsilon_{1} \hat{\chi}_{1} z^2_{2} + \zeta_1 r_1 |{z_{2}}| + {d}_{max2} |{z_{2}}| \\
&- \delta_1 \tilde{\chi}^2_1+\frac{1}{2} \epsilon_1 \left|z_{2}\right|^2 \tilde{\chi}_{1}
+  \delta_1 {\chi}^*_1 \tilde{\chi}_{1}
\end{aligned}
\end{equation} 
Assume that $\mu_1$ and $\lambda_1$ are any positive constants. Using Young's inequality, we can obtain:
\begin{equation}
\small
\begin{aligned}
\label{eq25}
\dot{V}_2 \leq& -{z_{1}} z_2  -\frac{1}{2} k_{1} z^2_{2}-\frac{1}{2}\epsilon_{1} {\chi}^*_{1} z^2_{2} + \frac{1}{4}\mu^{-1}_1 r^2_1 + \mu_1 \zeta^2_1 {z_{2}}^2   \\
&+\frac{1}{4} \lambda_1^{-1} + \lambda_1 {d}^2_{max2} {z_{2}}^2- \delta_1 \tilde{\chi}^2_1
-  \delta_1 {\chi}^*_1 \tilde{\chi}_{1}
\end{aligned}
\end{equation} 
Now, we can introduce a positive parameter, which the
adaptive is supposed to estimate, as:
\begin{equation}
\small
\begin{aligned}
\label{eq25}
\chi_1^* = \frac{2}{\epsilon_1}(\mu_1 \zeta^2_1+ \lambda_1 {d}^2_{max2})
\end{aligned}
\end{equation} 
From (41), we have:
\begin{equation}
\begin{aligned}
\label{eq25}
\dot{V}_2 \leq& -{z_{1}} z_2  - \Psi_2 V_2 + \frac{1}{4}\mu^{-1}_1 r_1 + \tilde{\mu}_2
\end{aligned}
\end{equation} 
where:
\begin{equation}
\small
\begin{aligned}
\label{eq25}
\Psi_2 = \min [k_1, \hspace{0.2cm} \delta_1 \gamma_1], \hspace{0.2cm} \tilde{\mu}_2 = \frac{1}{4} \lambda_1^{-1}+\frac{1}{2} \delta_1 {{\chi}^*_{1}}^2
\end{aligned}
\end{equation} 
Similarly, we can have another Lyapunov function, as:
\begin{equation}
\small
\begin{aligned}
\label{eq24}
{V}_{3} =\frac{1}{2} \hspace{0.1cm} [\alpha_{3}^{-1} s_{12} ^{-1}{z^2_{3}}+{\gamma^{-1}_{2}}\tilde{\chi}_{2}^2]
\end{aligned}
\end{equation} 
and if the actuator mechanism has a fourth subsystem:
\begin{equation}
\small
\begin{aligned}
\label{eq24}
{V}_{4} =\frac{1}{2} \hspace{0.1cm} [\alpha_{4}^{-1} s_{13} ^{-1}{z^2_{4}}+{\gamma^{-1}_{3}}\tilde{\chi}_{3}^2]
\end{aligned}
\end{equation} 
Similarly, if we define:
\begin{equation}
\small
\begin{aligned}
\label{eq35}
&\|\alpha_{3}^{-1} s_{12}^{-1} F_3\| \hspace{0.1cm} \leq \zeta_2 \hspace{0.1cm}r_2\hspace{0.1cm}, \hspace{0.2cm} \|\alpha_{3}^{-1} s_{12} ^{-1} {D_3}^*\| \hspace{0.1cm} \leq {d}_{max3}
\end{aligned}
\end{equation}
and:
\begin{equation}
\small
\begin{aligned}
\label{eq35}
&\|\alpha_{4}^{-1} s_{13}^{-1} F_4\| \hspace{0.1cm} \leq \zeta_3 \hspace{0.1cm}r_3\hspace{0.1cm}, \hspace{0.2cm} \|\alpha_{4}^{-1} s_{13} ^{-1} {D_4}^*\| \hspace{0.1cm} \leq {d}_{max4}
\end{aligned}
\end{equation}
we can also define:
\begin{equation}
\small
\begin{aligned}
\label{eq25}
&\chi_2^* = \frac{2}{\epsilon_2}(\mu_2 \zeta^2_2+ \lambda_2 {d}^2_{max3})\\
&\chi_3^* = \frac{2}{\epsilon_3}(\mu_3 \zeta^2_3+ \lambda_3 {d}^2_{max4})
\end{aligned}
\end{equation} 
In the same way, we can reach:
\begin{equation}
\small
\begin{aligned}
\label{eq25}
\dot{V}_3 \leq& - \Psi_3 V_3 + \frac{1}{4}\mu^{-1}_2 r_2 + \tilde{\mu}_3\\
\dot{V}_4 \leq& - \Psi_4 V_4 + \frac{1}{4}\mu^{-1}_3 r_3 + \tilde{\mu}_4
\end{aligned}
\end{equation} 
where:
\begin{equation}
\small
\begin{aligned}
\label{eq25}
\Psi_3 = \min [k_2, \hspace{0.2cm} \delta_2 \gamma_2], \hspace{0.2cm} \tilde{\mu}_3 = \frac{1}{4} \lambda_2^{-1}+\frac{1}{2} \delta_2 {{\chi}^*_{2}}^2\\
\Psi_4 = \min [k_3, \hspace{0.2cm} \delta_3 \gamma_3], \hspace{0.2cm} \tilde{\mu}_4 = \frac{1}{4} \lambda_3^{-1}+\frac{1}{2} \delta_3 {{\chi}^*_{3}}^2
\end{aligned}
\end{equation} 
Now, from (26), (41), (51), and (52), we can introduce a Lyapunov function for the whole system, as follows:
\begin{equation}
\small
\begin{aligned}
\label{eq25}
V= \sum_{k=0}^{\upsilon} V_{k+1}
\end{aligned}
\end{equation} 
where $\upsilon = 2$ or $3$ depending on the number of subsystems, which can be two ($k=0,...,2$: number of subsystems is three) or three ($k=0,...,3$: number of subsystems is four), and:
\begin{equation}
\small
\begin{aligned}
\label{eq25}
{V}_{k+1} =\frac{1}{2} \hspace{0.1cm} [\alpha_{k+1}^{-1} s_{1k} ^{-1}{z^2_{k+1}}+{\gamma^{-1}_{k}}\tilde{\chi}_{k}^2]
\end{aligned}
\end{equation}
We can define:
\begin{equation}
\small
\begin{aligned}
\label{eq25}
&S=\begin{bmatrix}
\alpha_{1}^{-1} & 0 & 0 & 0 \\
0 & \alpha_{2}^{-1} s_{11} ^{-1} & 0 & 0 \\
0 & 0 & \alpha_{3}^{-1} s_{12} ^{-1} & 0 \\
0 & 0 & 0 & \alpha_{4}^{-1} s_{13} ^{-1}
\end{bmatrix}\\
&x_e=\begin{bmatrix}
z_{1} \\
z_{2} \\
z_{3} \\
z_{4}
\end{bmatrix}, \hspace{0.2cm} \Gamma^{-1}=\begin{bmatrix}
\gamma^{-1}_{0} \\
\gamma^{-1}_{1} \\
\gamma^{-1}_{2} \\
\gamma^{-1}_{3}
\end{bmatrix}, \hspace{0.2cm} \tilde{X}=\begin{bmatrix}
\tilde{\chi}_{0} \\
\tilde{\chi}_{1} \\
\tilde{\chi}_{2} \\
\tilde{\chi}_{3}
\end{bmatrix}
\end{aligned}
\end{equation} 
Thus, from (59) and (60), we have:
\begin{equation}
\small
\begin{aligned}
\label{eq25}
V= \frac{1}{2} x_e^{\top} S x_e + \tilde{X}^{\top} \Gamma^{-1} \tilde{X}
\end{aligned}
\end{equation} 
From (39), (49), (56), and (61):
\begin{equation}
\small
\begin{aligned}
\label{eq25}
\dot{V} \leq& {z_{1}} z_2 -{z_{1}} z_2  - \Psi_1 V_1 + \frac{1}{4}\mu^{-1}_0 r_0 + \tilde{\mu}_1  - \Psi_2 V_2 \\
&+ \frac{1}{4}\mu^{-1}_1 r_1 + \tilde{\mu}_2- \Psi_3 V_3 + \frac{1}{4}\mu^{-1}_2 r_2 + \tilde{\mu}_3- \Psi_4 V_4\\
& + \frac{1}{4}\mu^{-1}_3 r_3 + \tilde{\mu}_4
\end{aligned}
\end{equation} 
By defining:
\begin{equation}
\small
\begin{aligned}
\label{eq25}
\Psi = \min [\Psi_1,...,\Psi_4], \hspace{0.2cm} \tilde{\mu} = \tilde{\mu}_1+\tilde{\mu}_2+\tilde{\mu}_3+\tilde{\mu}_4
\end{aligned}
\end{equation} 
we will have:
\begin{equation}
\small
\begin{aligned}
\label{eq25}
\dot{V} \leq-\Psi V+\frac{1}{4} \sum_{k=0}^{\upsilon+1} \mu_k^{-1} r_k^2+\tilde{\mu}
\end{aligned}
\end{equation} 
based on \cite{shahna2023exponential, shahna2024integrating, heydari2024robust}, we can obtain:
\begin{equation}
\small
\begin{aligned}
\label{905}
\|x_e\|^2 \leq \frac{\frac{2}{S_{min}}V\left(t_0\right) e^{-\bar{\iota}(t-t_0)}+\frac{2}{S_{min}} \tilde{\mu}\Psi^{-1}} {1-\overset{*}{Q}}
\end{aligned}
\end{equation}
It is significant that:
\begin{equation}
\small
\begin{aligned}
\label{906}
\sup _{t \in\left[t_0, \infty\right]}(\frac{\frac{2}{S_{min}}V(t_0) e^{-\bar{\iota}(t-t_0)}} {1-\overset{*}{Q}})\leq \frac{\frac{2}{S_{min}}V(t_0)} {1-\overset{*}{Q}}
\end{aligned}
\end{equation}
Thus, based on \cite{corless1993bounded} and \cite{heydari2024robust}, it is obvious from (65) that along with the adaptive laws provided, $x_e$ reaches a defined region $G_{0}\left(\bar{\tau}_0\right)$ in uniformly exponential convergence, such that:
\begin{equation}
\small
\begin{aligned}
\label{907}
G_{0}\left(\bar{\tau}_0\right):=\left\{\|x_e\| \leq \bar{\tau}_0 := \sqrt{\frac{\frac{2}{S_{min}} \tilde{\mu} \Psi^{-1}}{1-\overset{*}{Q}}}\right\}
\end{aligned}
\end{equation}
The radius of the region $G_0(\bar{\tau}_0)$ directly depends on the intensity of the uncertainties and disturbances.

\begin{figure}[h!]
    \hspace*{-0.0cm} % Adjust the value as needed
    \centering
    \scalebox{1.0}{\includegraphics[trim={0cm 0.0cm 0.0cm 0cm},clip,width=\columnwidth]{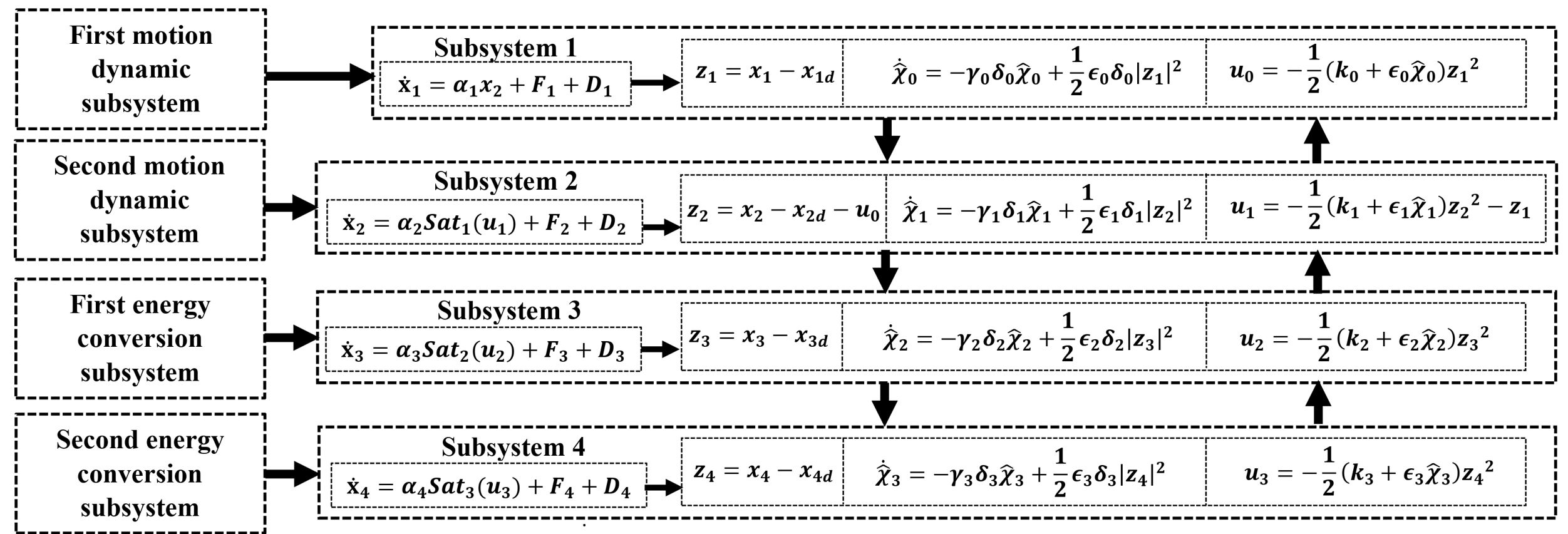}}
    \caption{Control schematic.}
    \label{fig7707}
\end{figure}

\section{Experimental Results OF THE Grc-Applied Actuation Mechanisms}
The GRC strategy was implemented in two experiments without relying on any information from the actuator system model.

\subsection{Experiment 1: GRC-applied EDA System}
\begin{figure}[h!]
    \hspace*{-0.0cm} % Adjust the value as needed
    \centering
    \scalebox{1}{\includegraphics[trim={0cm 0.0cm 0.0cm 0cm},clip,width=\columnwidth]{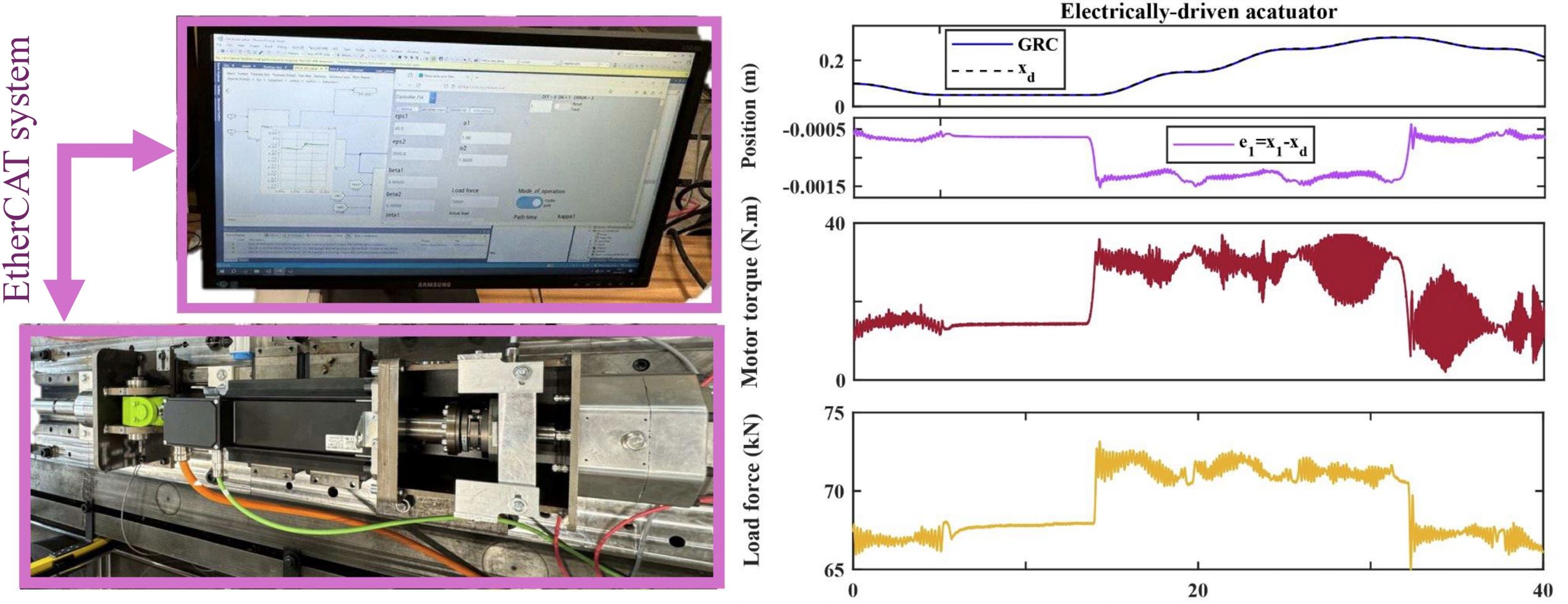}}
    \caption{The GRC-applied EDA performance with a $73$-kN load.}
    \label{ex1}
\end{figure}

This case study is a linear EDA powered by a PMSM, employing the GRC framework to track the linear desired position ${x}_d$. This setup was deployed on the Unidrive M700 controller, linked to an inverter that operates the servo motor with communication over an EtherCAT network. The GRC framework was configured with a sample time of 1,000 Hz. The GRC parameters were selected: $k_\upsilon = 35$, $\epsilon_\upsilon = 1$, $\gamma_\upsilon = 0.001$, and $\delta_\upsilon = 0.01$. In this experiment, the reference position trajectory $x_d$ was determined by quantic polynomials provided in Chapter 13 of \cite{jazar2010theory}. Fig. \ref{ex1} presents the results of this experiment, showing that the GRC strategy achieved approximately 1 mm accuracy in position tracking, within this complex linear EDA mechanism, with a $73$-kN load. This performance was over $200\%$ more accurate than PID control, and the position changes were significantly smoother than those observed with PID, validating GRC's robustness.

\subsection{Experiment 2: GRC-applied HDA system}
\begin{figure}[h!]
    \hspace*{-0.0cm} % Adjust the value as needed
    \centering
    \scalebox{1}{\includegraphics[trim={0cm 0.0cm 0.0cm 0cm},clip,width=\columnwidth]{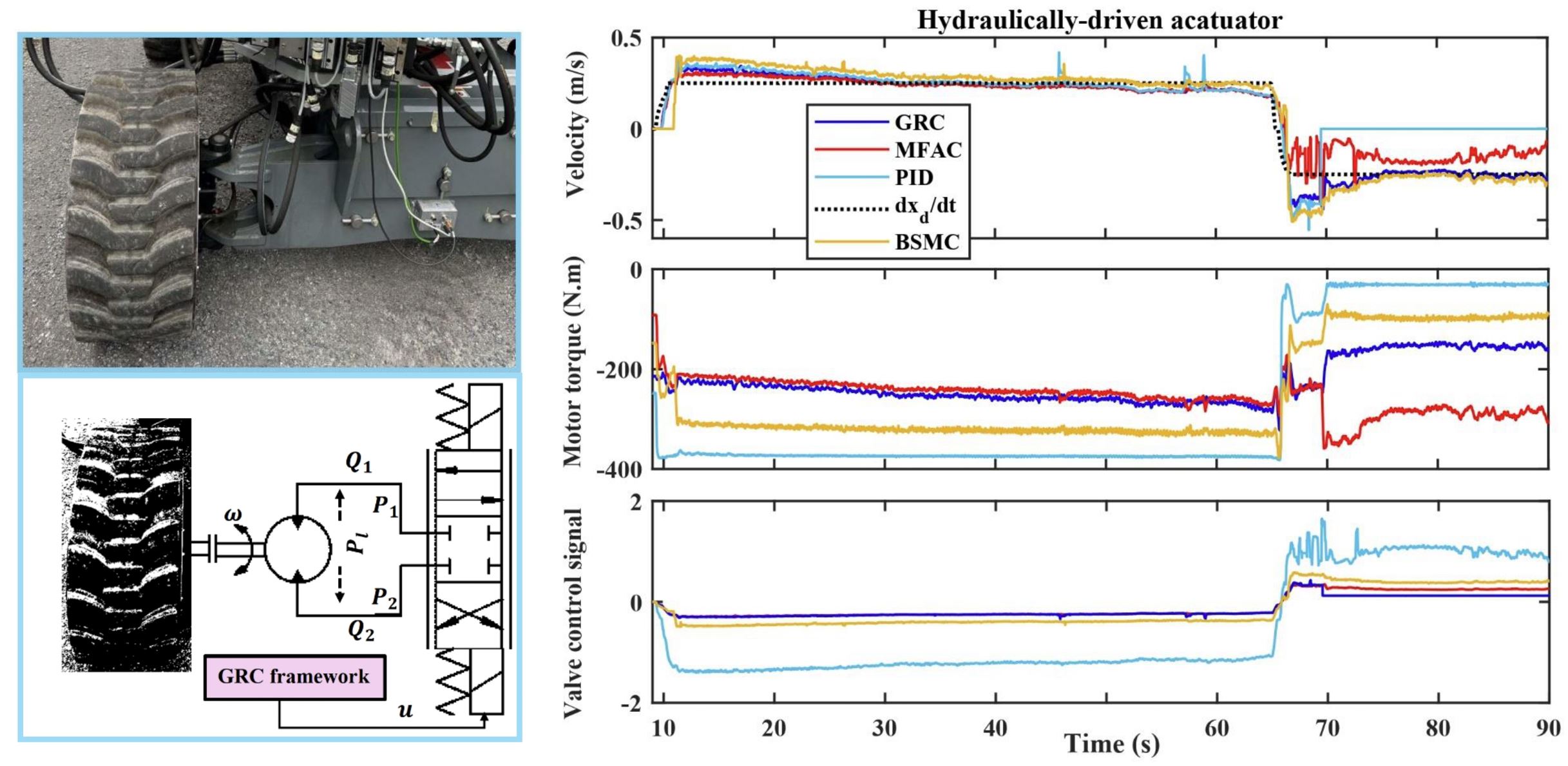}}
    \caption{The GRC-applied in-wheel HDA performance.}
    \label{ex2}
\end{figure}

This case study features a rotational valve-based HDA powered by a Bosch Rexroth pump, which rotates a hydraulic motor, specifically a Danfoss OMSS motor, to drive wheels with an $85$-cm radius and a $17.7$ gear ratio on a $5.5$-ton human-lifting mobile robot. This setup's central control unit is a Beckhoff IPC CX2030, operating in real time with a sampling rate of $1,000$ Hz. The GRC parameters were selected: $k_\upsilon = 3$, $\epsilon_\upsilon = 1$, $\gamma_\upsilon = 0.001$, and $\delta_\upsilon = 0.01$. In this test, a joystick determined the reference velocity $\frac{d x_d}{d t}$. Fig. \ref{ex2} demonstrates the performance of the GRC and three other model-free control strategies. It shows that PID, GRC, and model-free adaptive control (MFAC) \cite{wang2022research} exhibited faster transient responses in tracking the desired velocity trajectory. Meanwhile, the steady-state responses of GRC and backstepping sliding mode control (BSMC) \cite{truong2022backstepping} were superior to the others, achieving significantly more accurate tracking, highlighting their robust capabilities.
However, the proposed GRC slightly outperformed BSMC. In addition, although MFAC initially controlled the system effectively, its tracking accuracy diminished over time, while PID control lost effectiveness after a period. 

\section{Conclusion}
Despite significant advances, users still insist on utilizing common model-free controls, such as PID, due to their straightforward implementation. However, their effectiveness is limited in servo-driven actuator dynamics that exceed two orders and exhibit highly nonlinear interactions. This paper proposes a model-free GRC framework that can be straightforwardly applied to all complex servo-driven actuator systems, regardless of the power type of the servomechanisms, the control inputs of which were constrained according to the defined characteristics. The robustness and uniformly exponential stability in tracking desired motions of GRC-applied servo-driven actuation mechanisms were experimentally guaranteed by applying the GRC framework to two complex actuators.

\bibliographystyle{IEEEtran}
\bibliography{lcsys}

\begin{thebibliography}{10}
\providecommand{\url}[1]{#1}
\csname url@samestyle\endcsname
\providecommand{\newblock}{\relax}
\providecommand{\bibinfo}[2]{#2}
\providecommand{\BIBentrySTDinterwordspacing}{\spaceskip=0pt\relax}
\providecommand{\BIBentryALTinterwordstretchfactor}{4}
\providecommand{\BIBentryALTinterwordspacing}{\spaceskip=\fontdimen2\font plus
\BIBentryALTinterwordstretchfactor\fontdimen3\font minus \fontdimen4\font\relax}
\providecommand{\BIBforeignlanguage}[2]{{%
\expandafter\ifx\csname l@#1\endcsname\relax
\typeout{** WARNING: IEEEtran.bst: No hyphenation pattern has been}%
\typeout{** loaded for the language `#1'. Using the pattern for}%
\typeout{** the default language instead.}%
\else
\language=\csname l@#1\endcsname
\fi
#2}}
\providecommand{\BIBdecl}{\relax}
\BIBdecl

\bibitem{zhao2017pid}
C.~Zhao and L.~Guo, ``{}{PID} controller design for second order nonlinear uncertain systems,'' \emph{Sci. China Inf. Sci.}, vol.~60, pp. 1--13, 2017.

\bibitem{zhang2019theory}
J.~Zhang and L.~Guo, ``Theory and design of {}{PID} controller for nonlinear uncertain systems,'' \emph{IEEE L-CSS}, vol.~3, no.~3, pp. 643--648, 2019.

\bibitem{aastrom2006advanced}
K.~J. {\AA}str{\"o}m and T.~H{\"a}gglund, \emph{Advanced PID control}.\hskip 1em plus 0.5em minus 0.4em\relax ISA, 2006.

\bibitem{guo2017saturated}
Q.~Guo, J.~Yin, T.~Yu, and D.~Jiang, ``Saturated adaptive control of an electrohydraulic actuator with parametric uncertainty and load disturbance,'' \emph{IEEE Trans. Ind. Electron}, vol.~64, no.~10, pp. 7930--7941, 2017.

\bibitem{sciavicco2012modelling}
L.~Sciavicco and B.~Siciliano, \emph{Modelling and control of robot manipulators}.\hskip 1em plus 0.5em minus 0.4em\relax Springer Science \& Business Media, 2012.

\bibitem{mattila2017survey}
J.~Mattila, J.~Koivum{\"a}ki, D.~G. Caldwell, and C.~Semini, ``A survey on control of hydraulic robotic manipulators with projection to future trends,'' \emph{TMECH}, vol.~22, no.~2, pp. 669--680, 2017.

\bibitem{yao2017active}
J.~Yao and W.~Deng, ``Active disturbance rejection adaptive control of hydraulic servo systems,'' \emph{IEEE Trans. Ind. Electron.}, vol.~64, no.~10, pp. 8023--8032, 2017.

\bibitem{ren2019adaptive}
H.-P. Ren, X.~Wang, J.-T. Fan, and O.~Kaynak, ``Adaptive backstepping control of a pneumatic system with unknown model parameters and control direction,'' \emph{IEEE Acc.}, vol.~7, p. 64471–64482, 2019.

\bibitem{heydari2024robust}
M.~Heydari~Shahna, M.~Bahari, and J.~Mattila, ``Robust decomposed system control for an electro-mechanical linear actuator mechanism under input constraints,'' \emph{Int. J. Robust Nonlin.}, vol.~34, no.~7, pp. 4440--4470, 2024.

\bibitem{yang2017position}
X.~Yang, X.~Zheng, and Y.~Chen, ``Position tracking control law for an electro-hydraulic servo system based on backstepping and extended differentiator,'' \emph{TMECH}, vol.~23, no.~1, pp. 132--140, 2017.

\bibitem{yang2016disturbance}
J.~Yang, W.-H. Chen, S.~Li, L.~Guo, and Y.~Yan, ``Disturbance/uncertainty estimation and attenuation techniques in {}{PMSM} drives—a survey,'' \emph{IEEE Trans. Ind. Electron}, vol.~64, no.~4, pp. 3273--3285, 2016.

\bibitem{corradini2021robust}
M.~L. Corradini, ``A robust sliding-mode based data-driven model-free adaptive controller,'' \emph{IEEE L-CSS}, vol.~6, pp. 421--427, 2021.

\bibitem{truong2022backstepping}
H.-V.-A. Truong, H.-A. Trinh, and K.-K. Ahn, ``Backstepping sliding mode-based model-free control of electro-hydraulic systems,'' \emph{J. Driv. Control}, vol.~19, no.~1, pp. 51--61, 2022.

\bibitem{hamon2023model}
P.~Hamon, L.~Michel, F.~Plestan, and D.~Chablat, ``Model-free based control of a gripper actuated by pneumatic muscles,'' \emph{Mechatron.}, vol.~95, p. 103053, 2023.

\bibitem{wang2023continuous}
F.~Wang, Y.~Wei, H.~Young, D.~Ke, H.~Xie, and J.~Rodr{\'\i}guez, ``Continuous-control-set model-free predictive fundamental current control for {}{PMSM} system,'' \emph{IEEE Trans. Power Electron.}, vol.~38, no.~5, pp. 5928--5938, 2023.

\bibitem{jeong2022nonlinear}
Y.~W. Jeong and C.~C. Chung, ``Nonlinear proportional-integral disturbance observers for motion control of permanent magnet synchronous motors,'' \emph{IEEE L-CSS}, vol.~6, pp. 3062--3067, 2022.

\bibitem{abdeljawed2022simulation}
H.~B. Abdeljawed and L.~El~Amraoui, ``Simulation and rapid control prototyping of {}{DC} powered universal motors speed control: Towards an efficient operation in future {}{DC} homes,'' \emph{JESTECH}, vol.~34, p. 101092, 2022.

\bibitem{li2013adaptive}
Y.~Li, S.~Tong, and T.~Li, ``Adaptive fuzzy output feedback control for a single-link flexible robot manipulator driven {}{DC} motor via backstepping,'' \emph{Nonlinear Anal. Real World Appl.}, vol.~14, no.~1, pp. 483--494, 2013.

\bibitem{chatri2022design}
C.~Chatri, M.~Labbadi, M.~Ouassaid, K.~Elyaalaoui, and Y.~El~Houm, ``Design and implementation of finite-time control for speed tracking of permanent magnet synchronous motors,'' \emph{IEEE L-CSS}, vol.~7, pp. 721--726, 2022.

\bibitem{rezaeizadeh2024reliability}
A.~Rezaeizadeh and S.~Mastellone, ``Reliability and lifetime optimal control for electric vehicle power converters,'' \emph{IEEE L-CSS}, 2024.

\bibitem{wu2019robust}
S.~Wu, J.~Zhang, and B.~Chai, ``A robust backstepping sensorless control for interior permanent magnet synchronous motor using a super-twisting based torque observer,'' \emph{Asian J. Control}, vol.~21, no.~1, pp. 172--183, 2019.

\bibitem{manring2019hydraulic}
N.~D. Manring and R.~C. Fales, \emph{Hydraulic control systems}.\hskip 1em plus 0.5em minus 0.4em\relax John Wiley \& Sons, 2019.

\bibitem{wang2023position}
C.~Wang, Y.~Shi, Y.~Wang, S.~Xu, and Z.~Sun, ``Position tracking control for pneumatic servo system subject to state constraints and voltage saturation,'' \emph{TMECH}, 2023.

\bibitem{shahna2023exponential}
M.~H. Shahna and J.~Mattila, ``Exponential auto-tuning fault-tolerant control of n degrees-of-freedom manipulators subject to torque constraints,'' \emph{arXiv:2311.15852}, 2023.

\bibitem{shahna2024integrating}
M.~H. Shahna, S.~A.~A. Kolagar, and J.~Mattila, ``Integrating {}{D}eep{}{RL} with robust low-level control in robotic manipulators for non-repetitive reaching tasks,'' in \emph{IEEE ICMA}.\hskip 1em plus 0.5em minus 0.4em\relax IEEE, 2024, pp. 329--336.

\bibitem{corless1993bounded}
M.~Corless and G.~Leitmann, ``Bounded controllers for robust exponential convergence,'' \emph{J. Optim. Theory Appl.}, vol.~76, no.~1, pp. 1--12, 1993.

\bibitem{jazar2010theory}
R.~N. Jazar, \emph{Theory of applied robotics}.\hskip 1em plus 0.5em minus 0.4em\relax Springer, 2010.

\bibitem{wang2022research}
X.~Wang, Y.~Zhang, and C.~Li, ``Research on model-free adaptive control of electro-hydraulic servo system of continuous rotary motor,'' \emph{IEEE Acc.}, vol.~10, p. 31165–31174, 2022.

\end{thebibliography}

\end{document}